\documentclass[reqno,10pt,a4paper]{amsart}

\usepackage{amssymb}
\usepackage{url}

\allowdisplaybreaks[4]
\let\phi=\varphi
\let\kappa=\varkappa

\DeclareMathOperator{\sym}{sym}
\DeclareMathOperator{\const}{const}
\DeclareMathOperator{\Ei}{Ei}

\newcommand*{\abs}[1]{\left|#1\right|}

\theoremstyle{theorem}

\theoremstyle{definition}

\newtheorem*{examples}{Examples}
\theoremstyle{remark}

\usepackage[all]{xy}
\SelectTips{cm}{}

\usepackage{mathrsfs}
\let\mathcal\mathscr
\usepackage[mathscr]{eucal}

\newcommand{\cprime}{\/{\mathsurround=0pt$'$}}

\begin{document}

\title[Symmetry reductions and exact solutions]{Symmetry reductions and exact
  solutions of Lax integrable $3$-dimensional systems} \author{H.~Baran}
\address{Mathematical Institute, Silesian University in Opava, Na
  Rybn\'{\i}\v{c}ku 1, 746 01 Opava, Czech Republic}
\email{Hynek.Baran@math.slu.cz} \author{I.S.~Krasil{\cprime}shchik}
\address{Independent University of Moscow, B. Vlasevsky 11, 119002 Moscow,
  Russia \& Mathematical Institute, Silesian University in Opava, Na
  Rybn\'{\i}\v{c}ku 1, 746 01 Opava, Czech Republic}
\email{josephkra@gmail.com} \author{O.I.~Morozov} \address{Faculty of Applied
  Mathematics, AGH University of Science and Technology, Al. Mickiewicza 30,
  Krak\'ow 30-059, Poland} \email{morozov{\symbol{64}}agh.edu.pl}
\author{P.~Voj{\v{c}}{\'{a}}k} \address{Mathematical Institute, Silesian
  University in Opava, Na Rybn\'{\i}\v{c}ku 1, 746 01 Opava, Czech Republic}
\email{Petr.Vojcak@math.slu.cz}

\date{\today}

\begin{abstract}
  We present a complete description of $2$-dimensional equations that arise as
  symmetry reductions of fourf $3$-dimensional Lax-integrable equations: (1)
  the universal hierarchy equation~$u_{yy}=u_zu_{xy}-u_yu_{xz}$; (2) the 3D
  rdDym equation $u_{ty}=u_xu_{xy}-u_yu_{xx}$; (3) The basic Veronese web
  equation $u_{ty}=u_tu_{xy}-u_yu_{tx}$; (4) Pavlov's equation
  $u_{yy}=u_{tx}+u_yu_{xx}-u_xu_{xy}$.
\end{abstract}

\keywords{Partial differential equations, symmetry reductions, solutions}

\subjclass[2010]{35B06}

\maketitle

\tableofcontents
\newpage

\section*{Introduction}
\label{sec:introduction}

We consider the four $3$-dimensional Lax-integrable\footnote{We say that an
  equation is \emph{Lax-integrable} if it admits a zero-curvature
  representation with a non-removable parameter.} equations:
\begin{itemize}
\item the \emph{universal hierarchy equation}
  (Section~\ref{sec:univ-hier-equat})
  \begin{equation}\label{eq:34}
    u_{yy}=u_zu_{xy}-u_yu_{xz},
  \end{equation}
  see~\cite{UHE}.
\item the \emph{3D rdDym equation} (Section~\ref{sec:3d-rddym-equation})
  \begin{equation}\label{eq:35}
    u_{ty}=u_xu_{xy}-u_yu_{xx},
  \end{equation}
  see~\cite{3DrdDym,Ovs,Pav-1}.
\item the \emph{basic Veronese web equation}
  (Section~\ref{sec:basic-veronese-web})
\begin{equation}\label{eq:36}
  u_{ty}=u_tu_{xy}-u_yu_{tx},
\end{equation}
see~\cite{A-Sh,V-Web-1,F-Moss,V-Web-2}.
\item \emph{Pavlov's equation} (Section~\ref{sec:pavlovs-equation})
  \begin{equation}\label{eq:37}
  u_{yy}=u_{tx}+u_yu_{xx}-u_xu_{xy},
\end{equation}
see~\cite{Dun,Pav}.
\end{itemize}
Some of these equations arise also in~\cite{F-Kh} as integrable hydrodynamic
reductions of multi-dimensional dispersionless PDEs.

All the above listed equations may be obtained as the symmetry reductions of
the following Lax-integrable $4$-dimensional systems:
\begin{equation*}
u_{yz} = u_{tx} + u_x  u_{xy} - u_y  u_{xx}
\end{equation*}
and
\begin{equation*}
u_{ty} = u_z  u_{xy} - u_y  u_{xz}
\end{equation*}
introduced in~\cite{F-Kh} and~\cite{UHE}, respectively, while the latter two,
in turn, are the reductions of
\begin{equation*}
  u_{yz} = u_{ts} + u_s  u_{xz} - u_z  u_{xs}.
\end{equation*}
Here we give a complete answer to a natural question: what $2$-dimensional
equations are the reductions of the $3$-dimensional ones? The result
comprises~$32$ equations of which
\begin{itemize}
\item sixteen can be solved explicitly,
\item one reduces to the Riccati equation,
\item five can be linearized by the Legendre transformation,
\item while the rest ten are `nontrivial'.
\end{itemize}
The latter are presented in Table~\ref{tab:1} (in the third column, we
exemplify the simplest relations). The first two of these equations can be
transformed to the Liouville equation and the Gibbons-Tsarev equation,
respectively. The other eight, to our strong opinion, may possess interesting
integrability properties and we plan to study them in the nearest future.
\small\begin{table}[tbh] \centering
  \begin{align*}
    &\text{\underline{Reduction}}&&\text{\underline{of
        Eq.}}&&\text{\underline{Relations with the initial
        equation}}\\
    &2\Phi=\Phi\Phi_{xz}-\Phi_x\Phi_z,&&\eqref{eq:34}&&u=\frac{\Phi(x,z)}{y},\\
    &\Phi_{\xi\xi}=(\xi+\Phi_\xi)\Phi_{\eta\eta}-\Phi_\eta\Phi_{\xi\eta} - 2
    ,&&\eqref{eq:37}&&u=\Phi(\xi,\eta)+t^2\xi-2t\eta,\ \xi=y,\ \eta=x+ ty,\\
    &\Phi_{\xi\xi}=\Phi_x\Phi_\xi-\Phi\Phi_{x\xi},&&\eqref{eq:34}&&
    u=\Phi(x,\xi)e^{-z},\
    \xi=ye^{-z},\\
    &(1+\xi\Phi_z)\Phi_{\xi\xi}-\xi\Phi_\xi\Phi_{\xi
      z}+\Phi_\xi\Phi_z=0,&&\eqref{eq:34}&&
    u=\Phi(z,\xi)e^{-x},\ \xi=ye^{-x},\\
    &\Phi_\eta\Phi_{\xi\eta}-\Phi_\xi\Phi_{\eta\eta}=e^\eta\Phi_{\xi\xi},&&
    \eqref{eq:34}&&
    u=\Phi(\xi,\eta)e^{-x},\ \xi=ye^{-z},\ \eta=x-z,\\
    &(\xi+\Phi_\xi)\Phi_{\xi y}-\Phi_y(\Phi_{\xi\xi}+2)=0,&&
    \eqref{eq:35}&&u=\Phi(\xi,y)e^{2t},\
    \xi=xe^{t},\\
    &\Phi_{\xi t}=4\Phi\Phi_\xi-\xi\Phi_\xi^2+2\xi\Phi\Phi_{\xi\xi},
    &&\eqref{eq:35}&&u=\Phi(\xi,t)x^2,\ \xi=xe^{-y},\\
    &\Phi_{\eta\eta}+(\xi+\Phi_\eta)\Phi_{\xi\eta} = \Phi_\eta(2
    +\Phi_{\xi\xi}),&&\eqref{eq:35}&&u=\Phi(\xi,\eta)e^{2 t},\
    \xi=xe^{- t},\ \eta=y-t,\\
    &(4\xi^2-3\Phi)\Phi_{\xi\xi}-\Phi_{\xi t}-
    6\xi\Phi_\xi+\Phi_\xi^2+6\Phi=0,&&\eqref{eq:37}&& u=\Phi(\xi,y)y^3,\
    \xi=\frac{x}{y^2},\\
    &\Phi_{\xi\xi}=(\xi-\Phi_\eta)\Phi_{\xi\eta} +
    (2\eta+\Phi_\xi)\Phi_{\eta\eta} -\Phi_\eta=0,&&\eqref{eq:37}&&u =
    \Phi(\xi,\eta)e^{-3t},\ \xi=ye^{\beta t}, \eta=xe^{2 t}
\end{align*}
\caption{`Nontrivial' reductions}\label{tab:1}
\end{table}\normalsize
More detailed, but concise, infomation on the reductions may be also found in
Table~\ref{tab:summary}.

In Section~\ref{sec:preliminaries}, we briefly expose necessary preliminaries
(see, e.g.,~\cite{KLV}). In Section~\ref{sec:summary-results}, we present the
obtained results in a concise form.

\section{Preliminaries}
\label{sec:preliminaries}

Let~$\mathcal{E}$ be a differential equation given by
\begin{equation}
  \label{eq:29}
  F\Big(x,\dots,\frac{\partial^{\abs{\sigma}}u}{\partial x^\sigma},\dots\Big)=0,
\end{equation}
where~$u(x)$ is the unknown function in the variables~$x=(x^1,\dots,x^n)$. A
\emph{symmetry} of~$\mathcal{E}$ is a
function~$\phi=\phi(x,\dots,u_\sigma,\dots)$ in the \emph{jet
  variables}~$u_\sigma$, $\sigma$ being a multi-index, $u_{\varnothing}=u$,
that satisfies the \emph{linearized equation}
\begin{equation}
  \label{eq:30}
  \ell_{\mathcal{E}}(\phi)\equiv \sum_\sigma\frac{\partial F}{\partial
    u_\sigma}D_\sigma(\phi)=0,
\end{equation}
where~$D_\sigma=D_{i_1}\circ\dots\circ D_{i_k}$ for~$\sigma=i_1\dots i_k$, while
\begin{equation}
  \label{eq:31}
  D_i=\frac{\partial}{\partial x^i}+\sum_\sigma u_{\sigma
    i}\frac{\partial}{\partial u_\sigma}
\end{equation}
are the \emph{total derivatives} restricted to~$\mathcal{E}$. Symmetries
of~$\mathcal{E}$ form a Lie algebra~$\sym\mathcal{E}$ over~$\mathbb{R}$ with
respect to the \emph{Jacobi bracket}
\begin{equation}
  \label{eq:32}
  \{\phi,\bar{\phi}\}=\sum_\sigma\left(\frac{\partial\phi}{\partial
      u_\sigma}D_\sigma(\bar{\phi}) - \frac{\partial\bar{\phi}}{\partial
      u_\sigma}D_\sigma(\phi)\right).
\end{equation}

A solution~$u$ to Equation~\eqref{eq:29} is said to be \emph{invariant} with
respect to a symmetry~$\phi\in\sym\mathcal{E}$ if it enjoys the equation
\begin{equation}
  \label{eq:33}
  \phi\Big(x,\dots,\frac{\partial^{\abs{\sigma}}u}{\partial x^\sigma},\dots\Big)=0.
\end{equation}
The \emph{reduction} of~$\mathcal{E}$ with respect to~$\phi$ is
Equation~\eqref{eq:29} rewritten in terms of first integrals of
Equation~\eqref{eq:33}.

\section{The universal hierarchy equation}
\label{sec:univ-hier-equat}
The equation is
\begin{equation}
  \label{eq:1}
  u_{yy}=u_zu_{xy}-u_yu_{xz}.
\end{equation}

\subsection{Symmetries}
\label{sec:symmetries-1}
The defining equation for symmetries for~\eqref{eq:1} is
\begin{equation}
  \label{eq:5}
  D_y^2(\phi)=u_z D_xD_y(\phi)-u_yD_xD_z(\phi)+u_{xy}D_z(\phi)-u_{xz}D_y(\phi).
\end{equation}
Its solutions are
\begin{align*}
  \phi_1&=yu_y+u,\\
  \phi_2(X_2)&=X_2u_x-X_2'u,\\
  \phi_3(Z_3)&=Z_3u_z+Z_3'yu_y,\\
  \phi_4(Z_4)&=Z_4u_y,\\
  \phi_5(X_5)&=X_5,
\end{align*}
where~$X_i$ are functions of~$x$, $Z_i$ are functions of~$z$ and `prime'
denotes the derivative with respect to the corresponding variable. 
The commutator relations are given in Table~\ref{tab:symE_5}
\begin{table}[h]
  \centering
  \begin{tabular}{l|lllll}
    &$\phi_1$&$\phi_2(\bar{X}_2)$&$\phi_3{(\bar{Z}_3)}$&
    $\phi_4(\bar{Z}_4)$&$\phi_5(\bar{X}_5)$\\\hline 
    $\phi_1$&$0$&$0$&$0$&$\phi_4(\bar{Z}_4)$&$-\phi_5(\bar{X}_5)$\\
    $\phi_2(X_2)$&\dots&$\phi_2(\bar{X}_2X_2'-X_2\bar{X}_2')$&$0$&$0$&
    $\phi_5(\bar{X}_5X_2'-X_2\bar{X}_5')$\\ 
    $\phi_3(Z_3)$&\dots&\dots&$\phi_3(\bar{Y}_3Y_3'-Y_3\bar{Y}_3')$&
    $\phi_4(\bar{Y}_4Y_3'-Y_3\bar{Y}_4')$&$0$\\ 
    $\phi_4(Z_4)$&\dots&\dots&$0$&$0$&$0$\\
    $\phi_5(X_5)$&\dots&\dots&\dots&\dots&$0$
  \end{tabular}
  \caption{Lie algebra structure of $\sym\mathcal{E}_{\eqref{eq:1}}$}
  \label{tab:symE_5}
\end{table}

\subsection{Reductions}
\label{sec:reductions}

Thus, the general symmetry of Equation~\eqref{eq:1} is
\begin{equation*}
  \phi=X_2u_x+(\alpha y+Z_3'y+Z_4)u_y + Z_3u_z+ (\alpha-X_2')u+X_5,
\end{equation*}
where~$\alpha\in\mathbb{R}$ is a constant. Thus, invariant with respect
to~$\phi$ solutions are given by the system
\begin{equation}
  \label{eq:6}
  \frac{dx}{X_2}=\frac{dy}{(\alpha+Z_3')y+Z_4}= \frac{dz}{Z_3}
  =-\frac{du}{(\alpha-X_2')u+X_5}.
\end{equation}

We consider the following basic cases below:
\begin{description}
\item[Case~$00$] $X_2=0$, $Z_3=0$;
\item[Case~$01$] $X_2=0$, $Z_3\neq0$;
\item[Case~$10$] $X_2\neq0$, $Z_3=0$;
\item[Case~$11$] $X_2\neq0$, $Z_3\neq0$.
\end{description}

\subsubsection{Case~$00$}
\label{sec:case-00}

System~\eqref{eq:6} takes the form
\begin{equation*}
  \frac{dx}{0}=\frac{dy}{\alpha y+Z_4}= \frac{dz}{0}
  =-\frac{du}{\alpha u+X_5}.
\end{equation*}
Its integrals are
\begin{equation*}
  (\alpha y+Z_4)u+X_5y = \const,\quad x=\const,\quad z=\const
\end{equation*}
and the general solution is given by
\begin{equation*}
  \Psi((\alpha y+Z)u+Xy,x,z)=0,
\end{equation*}
where~$Z=Z_4$, $X=X_5$. Hence,
\begin{equation}
  \label{eq:12}
  u=\frac{\Phi(x,z)-Xy}{\alpha y+Z}.
\end{equation}
To simplify the subsequent exposition, we consider two subcases:
\begin{description}
\item[Subcase~$00.0$] $\alpha=0$;
\item[Subcase~$00.1$] $\alpha\neq0$.
\end{description}

\paragraph{\underline{Subcase~$00.0$}}
\label{sec:subcase-000}

After redenoting~$\Phi\mapsto\Phi/Z$, $Z\neq 0$, we have
\begin{equation}
  \label{eq:13}
  u=\Phi(x,z)-\frac{Xy}{Z}.
\end{equation}
Substituting to Equation~\eqref{eq:1}, one obtains
\begin{equation*}
  \frac{1}{Z}\cdot(X\Phi_{xz}-X'\Phi_z)=0,
\end{equation*}
which leads to the following class of solutions
\begin{equation*}\boxed{
  u=
  \begin{cases}
    \Phi(x,z),&\text{if }X=0,\\
    XP(z)+Q(x)-\frac{Xy}{Z},&\text{if }X\neq0.
  \end{cases}}
\end{equation*}

\paragraph{\underline{Subcase~$00.1$}}
\label{sec:subcase-001}
Making the change~$\Phi\mapsto\Phi-XZ$, one gets
\begin{equation*}
  u=\frac{\Phi}{y+Z}-X.
\end{equation*}
Substituting to~\eqref{eq:1}, one arrives to the equation
\begin{equation*}
  \boxed{2\Phi=\Phi\Phi_{xz}-\Phi_x\Phi_z.}
\end{equation*}
After the change~$\Phi=e^\Psi$ we obtain the Liouville equation
\begin{equation*}
  \boxed{\Psi_{xz}=2e^{-\Psi},}
\end{equation*}
see, e.g.~\cite{Liou}.

\subsubsection{Case~$01$}
\label{sec:case-01}
Now we have
\begin{equation*}
  \frac{dx}{0}=\frac{dy}{(\alpha+Z_3')y+Z_4}= \frac{dz}{Z_3}
  =-\frac{du}{\alpha u+X_5}.
\end{equation*}
The integrals of the system are
\begin{multline*}
  u\exp\left(\int\frac{\alpha dz}{Z_3}\right)+
  \int\frac{X_5}{Z_3}\exp\left(\int\frac{\alpha\, dz}{Z_3}\right)\,dz
  =\const,\quad 
  x=\const,\\ y\exp\left(-\int\frac{\alpha+Z_3'}{Z_3}\,dz\right) -
  \int\frac{Z_4}{Z_3}\exp\left(-\int\frac{\alpha+Z_3'}{Z_3}dz\right)\,dz
  =\const.  
\end{multline*}
We introduce new functions
\begin{equation*}
  Z=\int\frac{dz}{Z_3},\quad
  \bar{Z}=\int\frac{Z_4}{Z_3}
  \exp\left(-\int\frac{\alpha+Z_3'}{Z_3}\,dz\right)\,dz,\quad X=X_5.
\end{equation*}
Note that~$Z'\neq 0$, We and again distinguish two subcases:
\begin{description}
\item[Subcase~$01.0$] $\alpha=0$;
\item[Subcase~$01.1$] $\alpha\neq0$.
\end{description}

\paragraph{\underline{Subcase~$01.0$}}
\label{sec:subcase-010}
In this case, the system of integrals transforms to
\begin{equation*}
  u+XZ=\const,\quad x=\const,\quad yZ'-\bar{Z}=const,
\end{equation*}
and thus
\begin{equation}
  \label{eq:14}
  \Psi(u+XZ,x,yZ'-\bar{Z})=0
\end{equation}
is the general solution. Consequently,
\begin{equation*}
  u=\Phi(x,\xi)-XZ,
\end{equation*}
where~$\xi=yZ'-\bar{Z}$.  Substituting the last expression to
Equation~\eqref{eq:1}, we obtain the equation
\begin{equation}\label{eq:15}
  \boxed{\Phi_{\xi\xi}=X'\Phi_\xi-X\Phi_{x\xi}.}
\end{equation}
When~$X=0$, we obtain the solutions
\begin{equation*}
  \boxed{u=a_1(yZ'-\bar{Z})+a_0,\qquad a_i=a_i(x).}
\end{equation*}
If~$X\neq 0$ Equation~\eqref{eq:15} can also be solved and the general
solution is
\begin{equation*}
  \boxed{u=\Phi\left(yZ'-\bar{Z}+\int\frac{dx}{X}\right)+a(x),}
\end{equation*}
where~$\Phi$ is an arbitrary function in one argument.

\paragraph{\underline{Subcase~$01.1$}}
\label{sec:subcase-011}
We can set~$\alpha=1$ and the general solution is
\begin{equation*}
  \Psi((u+X)e^Z,x,Z'e^{-Z}y-\bar{Z})=0.
\end{equation*}
Hence, after the change~$Z\mapsto\ln Z$, we have
\begin{equation}
  \label{eq:16}
  u=\frac{1}{Z}\Phi(x,\xi)-X,
\end{equation}
where~$\xi=Z'y/Z^2-\bar{Z}$. Substituting~\eqref{eq:16} to
Equation~\eqref{eq:1}, one obtains
\begin{equation*}
  \boxed{\Phi_{\xi\xi}=\Phi_x\Phi_\xi-\Phi\Phi_{x\xi}.}
\end{equation*}

\subsubsection{Case~$10$}
\label{sec:case-10}
System~\eqref{eq:6} is now of the form
\begin{equation*}
   \frac{dx}{X_2}=\frac{dy}{\alpha y+Z_4}= \frac{dz}{0}
  =-\frac{du}{(\alpha-X_2')u+X_5}.
\end{equation*}
Then the integrals are
\begin{multline*}
  u\exp\left(\int\frac{\alpha-X_2'}{X_2}\,dx\right) +
  \int
  \frac{X_5}{X_2}\exp\left(\int\frac{\alpha-X_2'}{X_2}\,dx\right)\,dx=\const,\\
  y\exp\left(-\int\frac{\alpha\,dx}{X_2}\right)-
  \int\frac{Z_4}{X_2}\exp\left(-\int\frac{\alpha\,dx}{X_2}\right)\,dx= 
  \const,\ z=\const.
\end{multline*}
Let us introduce the notation
\begin{equation*}
  \int\frac{dx}{X_2}=X,\quad
  \frac{X_5}{X_2}\exp\left(\int\frac{\alpha-X_2'}{X_2}\,dx\right)\,dx=\bar{X},
  \quad Z_4=Z
\end{equation*}
and consider the subcases
\begin{description}
\item[Subcase~$10.0$] $\alpha=0$;
\item[Subcase~$10.1$] $\alpha\neq0$.
\end{description}

\paragraph{\underline{Subcase~$10.0$}}
\label{sec:subcase-100}
In this case the general solution is given by
\begin{equation*}
  \Psi(X'u+\bar{X},y-XZ,z)=0, \qquad X'\neq 0,
\end{equation*}
and thus
\begin{equation*}
  u=\frac{1}{X'}\Phi(\xi,z)-\bar{X},
\end{equation*}
where~$\xi=y-XZ$. Substituting this expression to Equation~\eqref{eq:1}, one
obtains
\begin{equation*}
  \boxed{(1+Z\Phi_z)\Phi_{\xi\xi}=Z\Phi_\xi\Phi_{\xi z}+Z'\Phi_\xi^2.}
\end{equation*}
The equation can be solved explicitly. Indeed, dividing by~$\Phi_\xi^2$ one
obtains
\begin{equation*}
  \frac{\Phi_{\xi\xi}}{\Phi_\xi^2}-Z'=Z\frac{\Phi_\xi\Phi_{\xi
      z}-\Phi_z\Phi_{\xi\xi}}{\Phi_\xi^2},
\end{equation*}
or
\begin{equation*}
  -\left(\frac{1}{\Phi_\xi}\right)_\xi-Z'=
  Z\left(\frac{\Phi_z}{\Phi_\xi}\right)_\xi.
\end{equation*}
Hence,
\begin{equation*}
  -\frac{1}{\Phi_\xi}-Z'\xi=Z\frac{\Phi_z}{\Phi_\xi}+\phi,
\end{equation*}
where~$\phi=\phi(z)$ is an arbitrary function. Thus,
\begin{equation*}
  Z\Phi_z+(Z'\xi+\phi)\Phi_\xi=-1
\end{equation*}
and
\begin{equation*}
  \boxed{\Phi=\Upsilon(\xi-\bar{\phi})-\int\frac{dz}{Z}}
\end{equation*}
is the general solution, where~$\bar{\phi}=Z\int\frac{\phi\,dz}{Z^2}$.

\paragraph{\underline{Subcase~$10.1$}}
\label{sec:subcase-101}
We may set~$\alpha=1$ and then obtain the general solution in the form
\begin{equation*}
  \Psi\left(X'ue^X+\bar{X},e^{-X}(y+Z),z\right)=0,
\end{equation*}
or, after the change~$X\mapsto\ln X$,
\begin{equation}\label{eq:17}
  u=\frac{\Phi(\xi,z)-\bar{X}}{X'},
\end{equation}
where~$\xi=(y+Z)/X$.
Substituting to~\eqref{eq:1}, one has
\begin{equation*}
  \boxed{(1+\xi\Phi_z)\Phi_{\xi\xi}-\xi\Phi_\xi\Phi_{\xi z}+\Phi_\xi\Phi_z=0.}
\end{equation*}

\subsubsection{Case~$11$}
\label{sec:case-11}
We have here
\begin{equation*}
  \frac{dx}{X_2}=\frac{dy}{(\alpha+Z_3')y+Z_4}= \frac{dz}{Z_3}
  =-\frac{du}{(\alpha-X_2')u+X_5},
\end{equation*}
where~$X_2\neq 0$, $Z_3\neq 0$. The integrals are
\begin{multline*}
  u\exp{\left(\int\frac{\alpha-X_2'}{X_2}\,dx\right)}+
  \int\frac{X_5}{X_2}\exp{\left(\int\frac{\alpha-X_2'}{X_2}\,dx\right)}\,dx
  =\const, \int\frac{dx}{X_2}-\int\frac{dx}{Z_3}
  =\const,\\
  y\exp{\left(-\int\frac{\alpha+Z_3'}{Z_3}\,dx\right)} -
  \int\frac{Z_4}{Z_3}\exp{\left(-\int\frac{\alpha+Z_3'}{Z_3}\,dx\right)}\,dx
  =\const.
\end{multline*}
As before, we introduce the notation
\begin{equation*}
  \int\frac{dx}{X_2}=X,\
  \int\frac{X_5}{X_2}\exp{\left(\int\frac{\alpha-X_2'}{X_2}\,dx\right)}\,dx 
  = \bar{X},\quad \int\frac{dx}{Z_3}=Z,\
  \int\frac{Z_4}{Z_3} \exp{\left(-\int\frac{\alpha+Z_3'}{Z_3}\,dx\right)}\,dx
  = \bar{Z} 
\end{equation*}
and consider two subcases
\begin{description}
\item[Subcase~$11.0$] $\alpha=0$;
\item[Subcase~$11.1$] $\alpha\neq0$.
\end{description}

\paragraph{\underline{Subcase~$11.0$}}
\label{sec:subcase-110}
The general solution is given by
\begin{equation*}
  \Psi(X'u+\bar{X},Z'y-\bar{Z}, X-Z)=0
\end{equation*}
in this case and thus
\begin{equation*}
  u=\frac{\Phi(\xi,\eta)-\bar{X}}{X'},\qquad\xi=Z'y-\bar{Z},\quad \eta=X-Z.
\end{equation*}
After substituting to Equation~\eqref{eq:1}, one has
\begin{equation*}
  \boxed{\Phi_{\xi\xi}=\Phi_\xi\Phi_{\eta\eta}-\Phi_\eta\Phi_{\xi\eta}.}
\end{equation*}
The equation can be linearized by the Legendre transformation,~\cite{MM}.

\paragraph{\underline{Subcase~$11.1$}}
\label{sec:subcase-111}
Setting~$\alpha=1$, we obtain the general solution
\begin{equation*}
  \Psi(X'e^Xu+\bar{X},Z'e^{-Z}y-\bar{Z},X-Z)=0
\end{equation*}
and, after the change~$X\mapsto\ln X$, $Z\mapsto-\ln Z$, one has
\begin{equation}\label{eq:18}
  u=\frac{\Phi(\xi,\eta)-\bar{X}}{X'},\qquad \xi=Z'y-\bar{Z},\quad
  \eta=\ln(XZ). 
\end{equation}
Substituting~\eqref{eq:18} to~\eqref{eq:1}, we obtain the equation
\begin{equation*}
  \boxed{\Phi_\eta\Phi_{\xi\eta}-\Phi_\xi\Phi_{\eta\eta}=e^\eta\Phi_{\xi\xi}.}
\end{equation*}

\section{The 3D rdDym equation}
\label{sec:3d-rddym-equation}

The equation is
\begin{equation}
  \label{eq:2}
  u_{ty}=u_xu_{xy}-u_yu_{xx}.
\end{equation}

\subsection{Symmetries}
\label{sec:symmetries}
Symmetries of Equation~\eqref{eq:2} are defined by
\begin{equation}
  \label{eq:7}
  D_tD_y(\phi)=u_xD_xD_y(\phi)-u_yD_x^2(\phi)+u_{xy}D_x(\phi)-u_{xx}D_y(\phi).
\end{equation}
Solutions of~\eqref{eq:7} are
\begin{align*}
  \phi_1&=xu_x-2u,\\
  \phi_2(T_2)&=T_2u_t+T_2'(xu_x-u)+\frac{1}{2}T_2''x^2,\\
  \phi_3(Y_3)&=Y_3u_y,\\
  \phi_4(T_4)&=T_4u_x+T_4'x,\\
  \phi_5(T_5)&=T_5,
\end{align*}
where~$Y_i=Y_i(y)$, $T_i=T_i(t)$ and `primes' denote the derivatives.
The commutator relations are given in Table~\ref{tab:symE_3}.
\begin{table}[h]
  \centering
  \begin{tabular}{l|lllll}
    &$\phi_1$&$\phi_2(\bar{T}_2)$&$\phi_3(\bar{Y}_3)$&$\phi_4(\bar{T}_4)$&
    $\phi_5(\bar{T}_5)$\\\hline 
    $\phi_1$&$0$&$0$&$0$&$\phi_4(\bar{T}_4)$&$2\phi_5(\bar{T}_5)$\\
    $\phi_2(T_2)$&\dots&$\phi_2(\bar{T}_2X_2'-T_2\bar{T}_2')$&$0$&
    $\phi_4(\bar{T}_4T_2'-T_2\bar{T}_4')$&
    $\phi_5(\bar{T}_5T_2'-T_2\bar{T}_5')$\\  
    $\phi_3(Y_3)$&\dots&\dots&$0$&$0$&$0$\\
    $\phi_4(T_4)$&\dots&\dots&\dots&
    $\phi_5(\bar{T}_4T_4'-T_4\bar{T}_4')$&$0$\\ 
    $\phi_5(T_5)$&\dots&\dots&\dots&\dots&$0$
  \end{tabular}
  \caption{Lie algebra structure of~$\sym\mathcal{E}_{\eqref{eq:2}}$}
  \label{tab:symE_3}
\end{table} 

\subsection{Reductions}
\label{sec:reductions-1}
The general symmetry of Equation~\eqref{eq:2} is
\begin{equation*}
  \phi=(\alpha x+T_2'x+T_4)u_x + Y_3u_y + T_2u_t - (2\alpha+T_2')u +T_5+T_4'x
  +\frac{1}{2}T_2''x^2,
\end{equation*}
where~$\alpha\in\mathbb{R}$ is a constant. Consequently, $\phi$-invariant
solutions are defined by the system
\begin{equation}
  \label{eq:8}
  \frac{dx}{(\alpha+T_2')x+T_4} = \frac{dy}{Y_3} = \frac{dt}{T_2} =
  \frac{du}{(2\alpha+T_2')u -T_5-T_4'x  -\frac{1}{2}T_2''x^2}.
\end{equation}
In what follows, we consider the following cases
\begin{description}
\item[Case~$00$] $Y_3=0$, $T_2=0$;
\item[Case~$01$] $Y_3=0$, $T_2\neq0$;
\item[Case~$10$] $Y_3\neq0$, $T_2=0$;
\item[Case~$11$] $Y_3\neq0$, $T_2\neq0$.
\end{description}

\subsubsection{Case~$00$}
\label{sec:case-00-1}
Equation~\eqref{eq:10} takes the form
\begin{equation*}
  \frac{dx}{\alpha x+T_4} = \frac{dy}{0} = \frac{dt}{0} =
  \frac{du}{2\alpha u -T_5-T_4'x}.
\end{equation*}

As before, two subcase must be considered:
\begin{description}
\item[Subcase~$00.0$] $\alpha=0$;
\item[Subcase~$00.1$] $\alpha\neq 0$.
\end{description}

\paragraph{\underline{Subcase~$00.0$}}
\label{sec:subcase-000-1}
Here we have
\begin{equation*}
  \frac{dx}{T_4} = \frac{dy}{0} = \frac{dt}{0} =
  -\frac{du}{T_5+T_4'x}
\end{equation*}
and the general solution is given by
\begin{equation*}
  \Psi\left(u+\frac{1}{2}Tx^2+\bar{T}x,y,t\right)=0,
\end{equation*}
or
\begin{equation}\label{eq:19}
  u=\Phi(y,t)-\frac{1}{2}Tx^2-\bar{T}x,
\end{equation}
where~$T=T_4'/T_4$, $\bar{T}=T_5/T_4$. Substituting~\eqref{eq:19} in
Equation~\eqref{eq:18}, we obtain
\begin{equation*}
  \boxed{\Phi_{yt}=T\Phi_y.}
\end{equation*}
The general solution is
\begin{equation*}
  \Phi=\phi(y)e^{\int T\,dt}+\psi(t),
\end{equation*}
which leads to the following family of solutions to Equation~\eqref{eq:18}:
\begin{equation*}
  \boxed{u=\phi(y)e^{\int T\,dt}+\psi(t)-\frac{1}{2}Tx^2-\bar{T}x.}
\end{equation*}

\paragraph{\underline{Subcase~$00.1$}}
\label{sec:subcase-001-1}
Setting~$\alpha=1$, we obtain
\begin{equation*}
  \frac{dx}{x+T_4} = \frac{dy}{0} = \frac{dt}{0} =
  \frac{du}{2 u -T_5-T_4'x}.
\end{equation*}
The general solution of this system is
\begin{equation}\label{eq:21}
  u=(x+T)^2\Phi(y,t)+T'(x+T)+\bar{T},
\end{equation}
where~$T=T_4$, $\bar{T}=(T_5-T_4T_4')/2$. Substituting to~\eqref{eq:2}, we
obtain the equation
\begin{equation*}
  \boxed{\Phi_{yt}=2\Phi\Phi_y.}
\end{equation*}
Integrating over~$y$, we come to the Riccati equation
\begin{equation*}
  \Phi_t=\Phi^2+\phi(t).
\end{equation*}
Thus, to any choice of~$\phi$ there corresponds a family of solutions to
Equation~\eqref{eq:2}.

\begin{examples}
  Let us consider some particular cases.
  \begin{enumerate}
  \item If~$\phi=0$ then
    \begin{equation*}
      \Phi = \frac{1}{\psi-t}.
    \end{equation*}
    Here and in all the examples below~$\psi$ is an arbitrary function
    of~$y$.
  \item For~$\phi=a^2$, $a=\const$, one has
    \begin{equation*}
      \Phi=a\tan\big(a(t+\psi)\big).
    \end{equation*}
  \item If~$\phi=-a^2$ then
    \begin{equation*}
      \Phi=a\frac{1+e^{2a(t+\psi)}}{1-e^{2a(t+2 \psi)}}
    \end{equation*}
  \item For~$\phi= t^{\kappa}$ one has
    \begin{equation*}
      \Phi = \frac{-\psi J(\frac{1}{\kappa+2}, \frac{2 t^{\frac{1}{2}
            \kappa+1}}{\kappa+2})
        +\psi J(\frac{3+\kappa}{\kappa+2}, \frac{2
          t^{\frac{1}{2}\kappa+1}}{\kappa+2}) t^{\frac{1}{2}\kappa+1}
        -Y(\frac{1}{\kappa+2}, \frac{2 t^{\frac{1}{2}\kappa+1}}{\kappa+2})
        +Y(\frac{3+\kappa}{\kappa+2}, 2
        \frac{t^{\frac{1}{2}\kappa+1}}{\kappa+2}) 
        t^{\frac{1}{2}\kappa+1}}
      {t \big(\psi J(\frac{1}{\kappa+2}, \frac{2 t^{\frac{1}{2}\kappa+1}}{\kappa+2})
        +Y(\frac{1}{\kappa+2}, \frac{2 t^{\frac{1}{2}\kappa+1)}}{\kappa+2})\big)},
    \end{equation*}
    where
    \begin{align*}
      J(a,\theta ) &= \sum_{m=0}^\infty \frac{(-1)^m}{m! \, \Gamma(m+a+1)}
      {\left(\frac{\theta }{2}\right)}^{2m+a},\\
      Y(a,\theta ) &= \frac{J(a,\theta ) \cos(a\pi) -
        J(-a,\theta )}{\sin(a\pi)}
    \end{align*}
    are Bessel  functions of the first and second kinds, respectively.
  \item If~$\phi= e^t$, then
    \begin{equation*}
      \Phi = \left(
        \frac{\psi Y(1,2 e^{\frac{t}{2}})}{\psi Y(0, 2 e^{\frac{t}{2}})
          +J(0, 2 e^{\frac{t}{2}})}
        +\frac{J(1, 2e^{\frac{t}{2}})}{\psi Y(0,2e^{\frac{t}{2}})
          +J(0, 2e^{\frac{t}{2}})          }
      \right) e^{\frac{t}{2}}
    \end{equation*}
  \item For~$\phi= (1-t)/(1+t)$ the solution is
    \begin{equation*}
      \Phi = \frac{2 \psi \Ei(1, -2-2 t) t+\psi e^{2+2 t}+2 t}
      {2 \psi (1+t) \Ei(1, -2-2 t)+2 t+\psi e^{2+2 t}+2},
    \end{equation*}
    where
    \begin{equation*}
      \Ei(a, t) = \int_1^\infty\frac{e^{-\theta t}\,d\theta}{\theta^a}
    \end{equation*}
    is the exponential integral function.
  \end{enumerate}
\end{examples}

\subsubsection{Case~$01$}
\label{sec:case-01-1}
We have
\begin{equation*}  
  \frac{dx}{(\alpha+T_2')x+T_4} = \frac{dy}{0} = \frac{dt}{T_2} =
  \frac{du}{(2\alpha+T_2')u -T_5-T_4'x  -\frac{1}{2}T_2''x^2},
\end{equation*}
where~$T_2\neq 0$. Its integrals are
\begin{multline*}
  T'xe^{-\alpha T}-\bar{T}=\const,\ y=\const,\\
  T'ue^{-2\alpha T}+\bar{\bar{T}} +
  \left(\alpha\bar{T}+\left(\frac{\bar{T}}{T'}\right)'\right)(T'xe^{-\alpha
    T}-\bar{T}) -\frac{1}{2}\frac{T''}{(T')^2}(T'xe^{-\alpha
    T}-\bar{T})^2=\const,
\end{multline*}
where
\begin{multline*}
  T=\int\frac{dt}{T_2},\quad
  \bar{T}=\int\left(T_4\cdot(T')^2\cdot e^{-\alpha T}\right)\,dt,\\
  \bar{\bar{T}}=\int\left(T_5\cdot(T')^2\cdot e^{-2\alpha T} + T_4'\cdot
    T\cdot T'\cdot e^{-\alpha T} + \frac{1}{2}T_2''\cdot\bar{T}^2\right)\,dt.
\end{multline*}
Then the general solution is
\begin{multline*}
  \Psi\left(T'xe^{-\alpha T}-\bar{T},y, T'ue^{-2\alpha}T+\bar{\bar{T}} +
    \left(\alpha\bar{T}+\left(\frac{\bar{T}}{T'}\right)'\right)(T'xe^{-\alpha
      T}-\bar{T})\right.\\
  \left.-\frac{1}{2}\frac{T''}{(T')^2}(T'xe^{-\alpha T}-\bar{T})^2\right)=0,
\end{multline*}
or
\begin{equation*}
  u=\left(\Phi(\xi,y)-\bar{\bar{T}} -\left(\alpha\bar{T} +
      \left(\frac{\bar{T}}{T'}\right)'\right)\xi
    +\frac{1}{2}\frac{T''}{(T')^2}\xi^2 \right)\frac{e^{2\alpha T}}{T'}, 
\end{equation*}
where
\begin{equation*}
  \xi=T'xe^{-\alpha T}-\bar{T}.
\end{equation*}
Substituting to Equation~\eqref{eq:2}, one obtains
\begin{equation}\label{eq:20}
  \boxed{(\alpha\xi + \Phi_\xi)\Phi_{\xi y}-\Phi_y(\Phi_{\xi\xi}+2\alpha)=0.}
\end{equation}

\subsubsection{Case~$10$}
\label{sec:case-10-1}
The defining equations are
\begin{equation}\label{eq:22}
  \frac{dx}{\alpha x+T_4}=\frac{dy}{Y_3}=\frac{dt}{0} = \frac{du}{2\alpha
    u-T_5 -T_4'x},
\end{equation}
where~$Y_3\neq 0$.
Below we consider the following subcases:
\begin{description}
\item[Subcase~$10.00$] $\alpha=0$, $T_4=0$;
\item[Subcase~$10.01$] $\alpha=0$, $T_4\neq0$;
\item[Subcase~$10.1$] $\alpha\neq0$
\end{description}
and introduce~$Y=Y(y)$ such that~$1/Y_3=Y'$.

\paragraph{\underline{Subcase~$10.00$}}
\label{sec:subcase-10.00}
In this case,~$T_5\neq 0$ and System~\eqref{eq:22} takes the form
\begin{equation*}
  \frac{dx}{0}=Y'\,dy=\frac{dt}{0}=-\frac{du}{T_5}.
\end{equation*}
Denote~$T_5=T$. Then the integrals are
\begin{equation*}
  x=\const,\quad t=\const,\quad u+YT=\const.
\end{equation*}
Then the general solution is given by
\begin{equation*}
  \Psi(u+YT,x,t)=0,
\end{equation*}
or
\begin{equation*}
  u=\Phi(x,t)-YT.
\end{equation*}
Substituting to Equation~\eqref{eq:2}, one obtains
\begin{equation*}
  -Y'T'=Y'T\Phi_{xx},
\end{equation*}
or, since~$Y'=1/Y_3\neq 0$,
\begin{equation*}
  \Phi_{xx}=-\frac{T'}{T}.
\end{equation*}
This delivers us the following family of solutions:
\begin{equation*}
  \boxed{u=-\frac{T'}{2T}x^2+\phi(t)x+\psi(t)-YT.}
\end{equation*}

\paragraph{\underline{Subcase~$10.01$}}
\label{sec:subcase-10.01}
The defining equations are now
\begin{equation*}
  \frac{dx}{T_4}=Y'\,dy=\frac{dt}{0}=-\frac{du}{T_5+T_4'x}.
\end{equation*}
Let us introduce the notation~$T_4=T$, $T_5/T_4=\bar{T}$. Then the integrals
are
\begin{equation*}
  x-YT=\const,\quad t=\const,\quad u+\frac{T'}{2T}x^2+\bar{T}x=\const
\end{equation*}
and the general solution is
\begin{equation*}
  \Psi(u+\frac{T'}{2T}x^2+\bar{T}x,x-YT,t)=0,
\end{equation*}
or
\begin{equation*}
  u=\Phi(\xi,t)-\frac{T'}{2T}x^2-\bar{T}x,
\end{equation*}
where~$\xi=x-YT$. Substituting to~\eqref{eq:2}, we obtain the linear equation
\begin{equation*}
  \boxed{\left(\frac{T'}{T}\xi+\bar{T} \right)\Phi_{\xi\xi}+\Phi_{\xi t}=0.}
\end{equation*}
The general solution of this equation is
\begin{equation*}
  \Phi=\phi(\eta)T+\psi(t),\qquad \eta=\frac{\xi}{T}-\int\frac{\bar{T}}{T}\,dt,
\end{equation*}
which gives the family of solutions to~\eqref{eq:2}:
\begin{equation*}
  \boxed{ u=\phi(\eta)T+\psi(t)-\frac{T'}{2T}x^2-\bar{T}x, \qquad
    \eta=\frac{\xi}{T}-\int\frac{\bar{T}}{T}\,dt.} 
\end{equation*}

\paragraph{\underline{Subcase~$10.1$}}
\label{sec:subcase-10.1}
We can assume~$\alpha=1$ and the defining equations become
\begin{equation*}
  \frac{dx}{x+T_4}=Y'\,dy=\frac{dt}{0} = \frac{du}{2u-T_5 -T_4'x}.
\end{equation*}
The integrals of this system are
\begin{equation*}
  (x+T)e^{-Y}=\const,\quad t=\const,\quad \frac{u-\bar{T}}{(x+T)^2} -
  \frac{T'}{x+T} =\const,
\end{equation*}
where~$T=T_4$, $\bar{T}=(T_5-T'T)/2$, and thus the general solution is
\begin{equation*}
  \Psi\left( \frac{u-\bar{T}}{(x+T)^2} - \frac{T'}{x+T}(x+T),e^{-Y},t\right)=0,
\end{equation*}
or
\begin{equation*}
  u=(x+T)^2\Phi(\xi,t)+T'(x+T)+\bar{T},\qquad \xi=(x+T)e^{-Y}.
\end{equation*}
Substituting to~\eqref{eq:2}, one obtains the equation
\begin{equation*}
  \boxed{\Phi_{\xi t}=4\Phi\Phi_\xi-\xi\Phi_\xi^2+2\xi\Phi\Phi_{\xi\xi}.}
\end{equation*}

\subsubsection{Case~$11$}
\label{sec:case-11-1}
Let us set~$Y'=1/Y_3\neq 0$ and~$T'=1/T_2\neq $. Then System~\eqref{eq:8}
becomes
\begin{equation*}
  \frac{dx}{(\alpha+T_2')x+T_4} = Y'\,dy = T'\,dt =
  \frac{du}{(2\alpha+T_2')u -T_5-T_4'x  -\frac{1}{2}T_2''x^2}.
\end{equation*}
The integrals are
\begin{multline*}
  T'xe^{-\alpha T}-\bar{T}=\const,\quad Y-T=\const,\\
  T'ue^{-2\alpha}+\bar{\bar{T}} +
  \left(\alpha\bar{T}+\left(\frac{\bar{T}}{T'}\right)'\right)(T'xe^{-\alpha
    T}-\bar{T}) -\frac{1}{2}\frac{T''}{(T')^2}(T'xe^{-\alpha
    T}-\bar{T})^2=\const,
\end{multline*}
where, as before,
\begin{multline*}
  T=\int\frac{dt}{T_2},\quad
  \bar{T}=\int\left(T_4\cdot(T')^2\cdot e^{-\alpha T}\right)\,dt,\\
  \bar{\bar{T}}=\int\left(T_5\cdot(T')^2\cdot e^{-2\alpha T} + T_4'\cdot
    T\cdot T'\cdot e^{-\alpha T} + \frac{1}{2}T_2''\cdot\bar{T}^2\right)\,dt.
\end{multline*}
Thus, the general solution is given by
\begin{multline*}
  \Psi\left(T'xe^{-\alpha T}-\bar{T},Y-T, T'ue^{-2\alpha}T+\bar{\bar{T}} +
    \left(\alpha\bar{T}+\left(\frac{\bar{T}}{T'}\right)'\right)(T'xe^{-\alpha
      T}-\bar{T})\right.\\
  \left.-\frac{1}{2}\frac{T''}{(T')^2}(T'xe^{-\alpha T}-\bar{T})^2\right)=0,
\end{multline*}
or
\begin{equation*}
  u=\left(\Phi(\xi,\eta)-\bar{\bar{T}} -\left(\alpha\bar{T} +
      \left(\frac{\bar{T}}{T'}\right)'\right)\xi
    +\frac{1}{2}\frac{T''}{(T')^2}\xi^2 \right)\frac{e^{2\alpha T}}{T'}, 
\end{equation*}
where
\begin{equation*}
  \xi=T'xe^{-\alpha T}-\bar{T},\qquad \eta=Y-T.
\end{equation*}
Substituting the last expression to Equation~\eqref{eq:2}, one obtains
\begin{equation*}
  \boxed{\Phi_{\eta\eta}+(\alpha\xi+\Phi_\eta)\Phi_{\xi\eta} =
    \Phi_\eta(2\alpha + \Phi_{\xi\xi}).}
\end{equation*}

\section{The basic Veronese web equation}
\label{sec:basic-veronese-web}

The equation is
\begin{equation}
  \label{eq:3}
  u_{ty}=u_tu_{xy}-u_yu_{tx}.
\end{equation}

\subsection{Symmetries}
\label{sec:symmetries-2}

Symmetries of~\eqref{eq:3} are defined by
\begin{equation}
  \label{eq:9}
  D_tD_y(\phi)=u_tD_xD_y(\phi)-u_yD_tD_x(\phi)+
  u_{xy}D_t(\phi)-u_{tx}D_y(\phi),
\end{equation}
whose solutions are
\begin{align*}
  \phi_1(T_1)&=T_1u_t,\\
  \phi_2(X_2)&=X_2u_x-X_2'u,\\
  \phi_3(Y_3)&=Y_3u_y,\\
  \phi_4(X_4)&=X_4,
\end{align*}
where~$X_i=X_i(x)$, $Y_i=Y_i(y)$, and~$T_i=T_i(t)$.
The commutator relations in~$\sym\mathcal{E}_{\eqref{eq:3}}$ are given in
Table~\ref{tab:symE_7}.
\begin{table}[h]
  \centering
  \begin{tabular}{l|llll}
    &$\phi_1(\bar{T_1})$&$\phi_2(\bar{X}_2)$&$\phi_3(\bar{Y}_3)$&
    $\phi_4(\bar{X}_4)$\\\hline 
    $\phi_1(T_1)$&$\phi_1(\bar{T}_1T_1'-T_1\bar{T}_1')$&$0$&$0$&$0$\\
    $\phi_2(X_2)$&\dots&$\phi_2(\bar{X}_2X_2'-X_2\bar{X}_2')$&$0$&
    $\phi_4(\bar{X}_4X_2'-X_2\bar{X}_4')$\\ 
    $\phi_3(Y_3)$&\dots&\dots&$\phi_3(\bar{Y}_3Y_3'-Y_3\bar{Y}_3')$&$0$\\
    $\phi_4(X_4)$&\dots&\dots&\dots&$0$
  \end{tabular}
  \caption{Lie algebra structure of $\sym\mathcal{E}_{\eqref{eq:3}}$}
  \label{tab:symE_7}
\end{table}

\subsection{Reductions}
\label{sec:reductions-2}

The general symmetry of Equation~\eqref{eq:3} is
\begin{equation*}
  \phi=X_2u_x+Y_3u_y+T_1u_t-X_2'u+X_4
\end{equation*}
and the corresponding invariant solutions must satisfy the system
\begin{equation}
  \label{eq:10}
  \frac{dx}{X_2}=\frac{dy}{Y_3}=\frac{dt}{T_1}=\frac{du}{X_2'u-X_4}.
\end{equation}

We consider below the following cases:
\begin{description}
\item[Case~$100$] $X_2\neq0$, $Y_3=0$, $Z_4=0$;
\item[Case~$010$] $X_2=0$, $Y_3\neq0$, $Z_4=0$;
\item[Case~$001$] $X_2=0$, $Y_3=0$, $Z_4\neq0$;
\item[Case~$011$] $X_2=0$, $Y_3\neq0$, $Z_4\neq0$; 
\item[Case~$101$] $X_2\neq0$, $Y_3=0$, $Z_4\neq0$;
\item[Case~$110$] $X_2\neq0$, $Y_3\neq0$, $Z_4=0$;
\item[Case~$111$] $X_2\neq0$, $Y_3\neq0$, $Z_4\neq0$;
\end{description}
and use the notation~$1/X_2=X'$, $1/Y_3=Y'$, $1/Z_4=Z'$ when it is well defined.

\subsubsection{Case~$100$}
\label{sec:case-100}
The defining equation is
\begin{equation*}
  \frac{dx}{X_2}=\frac{dy}{0}=\frac{dt}{0}=\frac{du}{X_2'u-X_4}.
\end{equation*}
The integrals are
\begin{equation*}
  Xu-\bar{X}=\const,\qquad y=\const,\qquad  t=\const,
\end{equation*}
and thus
\begin{equation*}
  \Psi(Xu-\bar{X},y,t)=0
\end{equation*}
is the general solution, where~$\bar{X}=\int X_4X'\,dx$. Consequently,
\begin{equation*}
  u=\frac{\Phi(y,t)+\bar{X}}{X}.
\end{equation*}
Substituting to~\eqref{eq:3}, one obtains
\begin{equation*}
  \boxed{\Phi_{yt}=0.}
\end{equation*}
Hence,~$\Phi=\phi(y)+\psi(t)$ and
\begin{equation*}
  \boxed{u=\frac{\phi(y)+\psi(t)+\bar{X}}{X}}
\end{equation*}
is a family of solutions to~\eqref{eq:3}.

\subsubsection{Case~$010$}
\label{sec:case-010}
The defining equation is
\begin{equation*}
  \frac{dx}{0}=\frac{dy}{Y_3}=\frac{dt}{0}=-\frac{du}{X_4}.
\end{equation*}
The integrals are
\begin{equation*}
  u+\bar{X}Y=\const,\quad x=\const,\quad t=\const,
\end{equation*}
where~$\bar{X}=X_4$. Then
\begin{equation*}
  \Psi(u+\bar{X}Y,x,t)=0
\end{equation*}
is the general solution and
\begin{equation*}
  u=\Phi(x,t)-\bar{X}Y.
\end{equation*}
Substituting to~\eqref{eq:3}, one obtains~$Y'(\bar{X}\Phi_{xt}-\bar{X}'\Phi_t)
=0$ and since~$Y'\neq 0$,
\begin{equation*}
  \boxed{\bar{X}\Phi_{xt}-\bar{X}'\Phi_t=0.}
\end{equation*}
Thus, if~$\bar{X}=0$ we obtain the obvious family of solutions
\begin{equation*}
  \boxed{u=\Phi(x,t).}
\end{equation*}
If~$\bar{X}\neq 0$ the corresponding family of solutions is
\begin{equation*}
  \boxed{u=\bar{X}\phi(t)+\psi(x)-\bar{X}Y.}
\end{equation*}

\subsubsection{Case~$001$}
\label{sec:case-001}
The defining equation is
\begin{equation*}
  \frac{dx}{0}=\frac{dy}{0}=\frac{dt}{T_1}=-\frac{du}{X_4}.
\end{equation*}
Then, again denoting~$\bar{X}=X_4$, we get the integrals
\begin{equation*}
  u+\bar{X}T=\const,\quad x=\const,\quad y=\const
\end{equation*}
and the general solution in the form
\begin{equation*}
  \Psi(u+\bar{X}T,x,y)=0,
\end{equation*}
or
\begin{equation*}
  u=\Phi(x,y)-\bar{X}T.
\end{equation*}
Substituting to~\eqref{eq:3}, one obtains
\begin{equation*}
  \boxed{\bar{X}\Phi_{xy}-\bar{X}'\Phi_y=0,}
\end{equation*}
since~$T'\neq 0$. Then in the case~$\bar{X}=0$ we get the family of solutions
\begin{equation*}
  \boxed{u=\Phi(x,y)}
\end{equation*}
and when~$\bar{X}\neq 0$ the family
\begin{equation*}
  \boxed{u=\bar{X}\phi(y)+\psi(x)-\bar{X}T.}
\end{equation*}

\subsubsection{Case~$011$}
\label{sec:case-011}
The defining equation is
\begin{equation*}
  \frac{dx}{0}=\frac{dy}{Y_3}=\frac{dt}{T_1}=-\frac{du}{X_4}.
\end{equation*}
Its integrals are
\begin{equation*}
  x=\const,\quad Y-T=\const,\quad u+\bar{X}Y=\const
\end{equation*}
and the general solution is
\begin{equation*}
  \Psi(u+\bar{X}Y,x,Y-T)=0,
\end{equation*}
or
\begin{equation*}
  u=\Phi(x,\xi)-\bar{X}Y,\qquad \xi=Y-T.
\end{equation*}
Substituting to Equation~\eqref{eq:3}, one obtains
\begin{equation*}
  \boxed{\Phi_{\xi\xi}=\bar{X}\Phi_{x\xi}-\bar{X}'\Phi_\xi.}
\end{equation*}
If~$\bar{X}=0$ then
\begin{equation*}
  \boxed{u=\phi(x)+\psi(Y-T)-\bar{X}Y.}
\end{equation*}
In the case~$\bar{X}\neq 0$ the corresponding family is
\begin{equation*}
  \boxed{u=\bar{X}\phi\left(Y-T+\int\frac{dx}{\bar{X}}\right)+
    \psi(x)-\bar{X}Y.} 
\end{equation*}

\subsubsection{Case~$101$}
\label{sec:case-101}
The defining equation is
\begin{equation*}
  \frac{dx}{X_2}=\frac{dy}{0}=\frac{dt}{T_1}=\frac{du}{X_2'u-X_4}.
\end{equation*}
The integrals of this system are
\begin{equation*}
  X'u+\bar{X}=\const,\quad X-T=\const,\quad y=\const,
\end{equation*}
where~$\bar{X}=\int(X')^2X_4\,dx$ and the general solution is given by
\begin{equation*}
  \Psi(X'u-\bar{X},X-T,y)=0,
\end{equation*}
or
\begin{equation*}
  u=\frac{\Phi(y,\xi)+\bar{X}}{X'},\qquad \xi=X-T.
\end{equation*}
After substitution to Equation~\eqref{eq:3} one obtains
\begin{equation*}
  \boxed{(1+\Phi_\xi)\Phi_{y\xi}=\Phi_y\Phi_{\xi\xi}.}
\end{equation*}
The general solution to this equation is
\begin{equation*}
  \Psi=\Upsilon\left(\xi+\psi(y)\right)-\xi
\end{equation*}
and thus we get the family
\begin{equation*}
  \boxed{u=\frac{\Upsilon\big(X-T+\psi(y)\big)-X+T+\bar{X}}{X'}}
\end{equation*}
to Equation~\eqref{eq:3}.

\subsubsection{Case~$110$}
\label{sec:case-110}
The defining equation is
\begin{equation*}
  \frac{dx}{X_2}=\frac{dy}{Y_3}=\frac{dt}{0}=\frac{du}{X_2'u-X_4}.
\end{equation*}
The integrals of this system are
\begin{equation*}
  X'u-\bar{X}=\const,\quad X-Y=\const,\quad y=\const,
\end{equation*}
where~$\bar{X}=\int(X')^2X_4\,dx$ and the general solution is given by
\begin{equation*}
  \Psi(X'u+\bar{X},X-Y,y)=0,
\end{equation*}
or
\begin{equation*}
  u=\frac{\Phi(y,\xi)+\bar{X}}{X'},\qquad \xi=X-Y.
\end{equation*}
Substitution to Equation~\eqref{eq:3} leads to
\begin{equation*}
  \boxed{(1+\Phi_\xi)\Phi_{t\xi}=\Phi_t\Phi_{\xi\xi}.}
\end{equation*}
Similar to the previous case, we solve this equation and obtain the following
family of solutions to Equation~\eqref{eq:3}:
\begin{equation*}
  \boxed{u=\frac{\Upsilon\big(X-Y+\psi(t)\big)-X+Y+\bar{X}}{X'}.}
\end{equation*}

\subsubsection{Case~$111$}
\label{sec:case-111}
The defining equation is
\begin{equation*}
  \frac{dx}{X_2}=\frac{dy}{Y_3}=\frac{dt}{T_1}=\frac{du}{X_2'u-X_4}.
\end{equation*}
The integrals are
\begin{equation*}
  X'u-\bar{X}=\const,\quad X-Y=\const,\quad X-T=\const,
\end{equation*}
where, as before~$\bar{X}=\int(X')^2X_4\,dx$. This delivers the general
solution
\begin{equation*}
  \Psi(X'u-\bar{X},X-Y,X-T)=0,
\end{equation*}
i.e.,
\begin{equation*}
  u=\frac{\Phi(\xi,\eta)+\bar{X}}{X'},\qquad\xi=X-Y,\quad\eta=X-T.
\end{equation*}
Substituting to~\eqref{eq:3}, we obtain the equation
\begin{equation}\label{eq:28}
  \boxed{\Phi_\eta\Phi_{\xi\xi}+\left(\Phi_\eta-\Phi_\xi-1\right)\Phi_{\xi\eta}
    -\Phi_\xi\Phi_{\eta\eta}=0.}
\end{equation}
The equation linearizes by the Legendre transformation,~\cite{MM}.

\section{Pavlov's equation}
\label{sec:pavlovs-equation}
The equation is
\begin{equation}
  \label{eq:4}
  u_{yy}=u_{tx}+u_yu_{xx}-u_xu_{xy}.
\end{equation}

\subsection{Symmetries}
\label{sec:symmetries-3}
The defining equation for symmetries of~\eqref{eq:4} is
\begin{equation}
  \label{eq:11}
  D_y^2(\phi)=D_tD_x(\phi)+u_yD_x^2(\phi)-u_xD_xD_y(\phi)+u_{xx}D_y(\phi)
  -u_{xy} D_x(\phi).
\end{equation}
Its solutions are
\begin{align*}
  \phi_1&=2x-yu_x,\\
  \phi_2&=3u-2xu_x-yu_y,\\
  \phi_3(T_3)&=T_3u_t+T_3'(xu_x+yu_y-u)+\frac{1}{2}T_3''(y^2u_x-2xy)-
  \frac{1}{6}T_3'''y^3,\\ 
  \phi_4(T_4)&=T_4u_x-T_4'y,\\
  \phi_5(T_5)&=T_5u_y+T_5'(yu_x-x)-\frac{1}{2}T_5''y^2,\\
  \phi_6(T_6)&=T_6,
\end{align*}
where~$T_i$ are functions of~$t$. 
The Lie algebra structure in~$\sym\mathcal{E}_{\eqref{eq:4}}$ is given in
Table~\ref{tab:symE_12}.
\begin{table}[h]
  \centering
  \begin{tabular}{l|llllll}
    &$\phi_1$&$\phi_2$&$\phi_3(\bar{T}_3)$&$\phi_4(\bar{T}_4)$&
    $\phi_5(\bar{T}_5)$&$\phi_6(\bar{T}_6)$\\\hline 
    $\phi_1$&$0$&$\phi_1$&$0$&$2\phi_6(\bar{T}_4)$&
    $-2\phi_4(\bar{T}_5)$&$0$\\ 
    $\phi_2$&\dots&$0$&$0$&$-2\phi_4(\bar{T}_4)$&$-\phi_5(\bar{T}_5)$&
    $-3\phi_6(\bar{T}_6)$\\ 
    $\phi_3(T_3)$&\dots&\dots&$\phi_3(\bar{T}_3T_3'-T_3\bar{T}_3')$&
    $\phi_4(\bar{T}_4T_3'-T_3\bar{T}_4')$&
    $\phi_5(\bar{T}_5T_3'-T_3\bar{T}_5')$&
    $\phi_6(\bar{T}_6T_3-T_3\bar{T}_6')$\\   
    $\phi_4(T_4)$&\dots&\dots&\dots&$0$&
    $\phi_6(\bar{T}_5T_4'-T_4\bar{T}_5')$&$0$\\ 
    $\phi_5(T_5)$&\dots&\dots&\dots&\dots&
    $\phi_4(\bar{T}_5T_5'-T_5\bar{T}_5')$&$0$\\ 
    $\phi_6(T_6)$&\dots&\dots&\dots&\dots&\dots&$0$
  \end{tabular}
  \caption{Lie algebra structure of~$\sym\mathcal{E}_{\eqref{eq:4}}$}
  \label{tab:symE_12}
\end{table}

\subsection{Reductions}
\label{sec:reductions-3}
The general symmetry of Equation~\eqref{eq:4} is
\begin{multline*}
  \phi=\left(-\alpha y-2\beta x+T_3'x +\frac{1}{2}T_3''y^2 + T_4+
    T_5'y\right)u_x + (-\beta y+T_3'y+T_5)u_y + T_3u_t \\+(3\beta - T_3')u
  +2\alpha x - T_3''xy - \frac{1}{6}T_3'''y^3 - T_5'x -\frac{1}{2}T_5''y^2 +
  T_6 -T_4'y,
\end{multline*}
where~$\alpha$, $\beta\in\mathbb{R}$ are constants. Then the $\phi$-invariant
solutions are determined by the system
\begin{multline}
  \frac{dx}{(T_3'-2\beta) x + (T_5'-\alpha) y +\frac{1}{2}T_3''y^2 + T_4}
  =\frac{dy}{(T_3'-\beta) y+T_5} =\frac{dt}{T_3}\\= \frac{du}{(T_3' - 3\beta)u
    + (T_5'-2\alpha) x + T_4'y+ T_3''xy + \frac{1}{2}T_5''y^2 +
    \frac{1}{6}T_3'''y^3 - T_6}.
\end{multline}
We consider the following tree of options:
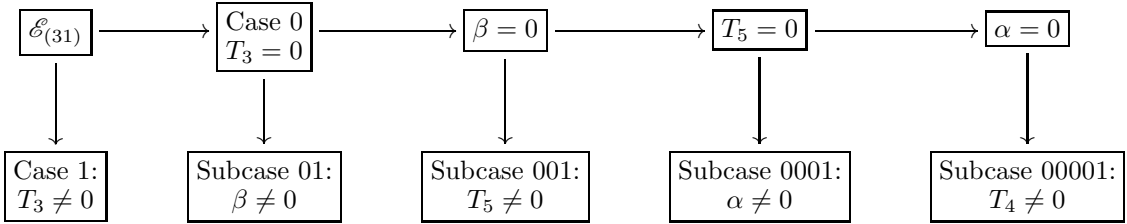
\begin{equation*}
  \xymatrix{
    \boxed{\mathcal{E}_{\eqref{eq:4}}}\ar[r]\ar[d]
    &\boxed{\txt{Case~$0$\\$T_3=0$}}\ar[r]\ar[d]
    &\boxed{\beta=0}\ar[r]\ar[d]&\boxed{T_5=0}\ar[d]\ar[r]
    &\boxed{\alpha=0}\ar[d]
    \\
    \boxed{\txt{Case~$1$:\\$T_3\neq0$}}
    &\boxed{\txt{Subcase~$01$:\\$\beta\neq0$}}
    &\boxed{\txt{Subcase~$001$:\\$T_5\neq0$}}
    &\boxed{\txt{Subcase~$0001$:\\$\alpha\neq0$}}
    &\boxed{\txt{Subcase~$00001$:\\$T_4\neq0$}}
  }
\end{equation*}

\subsubsection{Case~$0$}
\label{sec:case-0}
The defining equations are
\begin{equation*}
  \frac{dx}{-2\beta x + (T_5'-\alpha) y  + T_4}
  =\frac{dy}{-\beta y+T_5} =\frac{dt}{0}= \frac{du}{ - 3\beta u
    + (T_5'-2\alpha) x + T_4'y+  \frac{1}{2}T_5''y^2  - T_6}.
\end{equation*}
Due to the above picture, consider the following subcases:
\begin{description}
\item[Subcase~$00001$] $\beta=0$, $T_5=0$, $\alpha=0$, $T_4\neq0$;
\item[Subcase~$0001$] $\beta=0$, $T_5=0$, $\alpha\neq0$;
\item[Subcase~$001$] $\beta=0$, $T_5\neq0$;
\item[Subcase~$01$] $\beta\neq0$.
\end{description}

\paragraph{\underline{Subcase~$00001$}}
\label{sec:subcase-0001}
The defining equations are
\begin{equation*}
  \frac{dx}{T_4} =\frac{dy}{0} =\frac{dt}{0}= \frac{du}{T_4'y - T_6}.
\end{equation*}
Then the integrals are
\begin{equation*}
  u-(Ty-\bar{T})x=\const,\qquad y=\const,\qquad t=\const,
\end{equation*}
where~$T=T_4'/T_4$, $\bar{T}=T_6/T_4$. Thus,
\begin{equation*}
  \Psi(u-(Ty-\bar{T})x,y,t)=0
\end{equation*}
is the general solution and
\begin{equation*}
  u=\Phi(y,t)+(Ty-\bar{T})x.
\end{equation*}
Substituting to Equation~\eqref{eq:4}, one obtains
\begin{equation*}
  \Phi_{yy}=(T'-T^2)y+T\bar{T}-\bar{T}',
\end{equation*}
which gives the family
\begin{equation*}
  \boxed{u=\frac{1}{6}(T'-T^2)y^3+\frac{1}{2}(T\bar{T}-\bar{T}')y^2 + \phi(t)y
    +\psi(t) + (Ty-\bar{T})x.}
\end{equation*}
of exact solutions to~\eqref{eq:4}.

\paragraph{\underline{Subcase~$0001$}}
\label{sec:underl-0001}
We may assume~$\alpha=-1$ and the defining equations become
\begin{equation*}
  \frac{dx}{y  + T_4}
  =\frac{dy}{0} =\frac{dt}{0}= \frac{du}{2x + T_4'y  - T_6}. 
\end{equation*}
Then the integrals are
\begin{equation*}
  (y+T_4)u-x^2-(T_4'y-T_6)x=\const,\quad y=\const,\quad t=\const.
\end{equation*}
Consequently, the general solution is given by
\begin{equation*}
  \Psi\left((y+T_4)u-x^2-(T_4'y-T_6)x,y,t\right)=0
\end{equation*}
and thus
\begin{equation}\label{eq:23}
  u=\frac{\Phi(y,t)+x^2+(T_4'y-T_6)x}{y+T_4},
\end{equation}
or
\begin{equation*}
  u=\Phi(y,t)+T'x+\frac{x^2-\bar{T}x}{y+T},
\end{equation*}
where~$T=T_4$, $\bar{T}=T_4T_4'+T_6$. After substituting to~\eqref{eq:4}, we
obtain the equation
\begin{equation*}
  \boxed{\Phi_{yy}=\frac{2\Phi_y}{y+T}+T''-\frac{\bar{T}'}{y+T} +
    \frac{\bar{T}^2}{(y+T)^3}.}
\end{equation*}
Solving this equation, we obtain the following family of solutions to
Equation~\eqref{eq:4}:
\begin{equation*}
  \boxed{u=\phi(t)(y+T)^3-\frac{1}{2}T''(y+T)^2+\frac{1}{2}\bar{T}'(y+T) +
    \frac{2x^2 - 2\bar{T}x +\bar{T}^2}{y+T}+T'x+\psi(t).}
\end{equation*}

\paragraph{\underline{Subcase~$001$}}
\label{sec:subcase-001-2}
The defining equations are
\begin{equation*}
  \frac{dx}{(T_5'-\alpha) y  + T_4}
  =\frac{dy}{T_5} =\frac{dt}{0}= \frac{du}{(T_5'-2\alpha) x + T_4'y+
    \frac{1}{2}T_5''y^2  - T_6}.
\end{equation*}
Let us introduce the notation~$T=1/T_5$, $\bar{T}=T_4/T_5$,
$\bar{\bar{T}}=T_6/T_5$. Then the integrals acquire the form
\begin{multline*}
  t=\const,\quad x+\frac{1}{2}\left(\frac{T'}{T}+\alpha T\right)y^2 - \bar{T}y
  =\const,\\
  u+\left(\frac{1}{6}\frac{T''}{T} + \alpha T' +
    \frac{2}{3}\alpha^2T^2\right)y^3 - \frac{1}{2}\left(\bar{T}'+2\alpha
    T\bar{T}\right)y^2 +  
  \left(\left(\frac{T'}{T}+2\alpha T\right)x+\bar{\bar{T}}\right)y =\const.
\end{multline*}
Hence, the general solution is
\begin{multline*}
  \Psi\left(u+\left(\frac{1}{6}\frac{T''}{T} + \alpha T' +
    \frac{2}{3}\alpha^2T^2\right)y^3 - \frac{1}{2}\left(\bar{T}'+2\alpha
    T\bar{T}\right)y^2 +  
  \left(\left(\frac{T'}{T}+2\alpha T\right)x+\bar{\bar{T}}\right)y,\right.\\
  \left. x+\frac{1}{2}\left(\frac{T'}{T}+\alpha T\right)y^2 -
    \bar{T}y,t\right)=0,
\end{multline*}
or
\begin{equation*}
  u=\Phi(\xi,t)
  -\left(\frac{1}{6}\frac{T''}{T} + \alpha T' +
    \frac{2}{3}\alpha^2T^2\right)y^3
  + \frac{1}{2}\left(\bar{T}'+2\alpha
    T\bar{T}\right)y^2
  -  \left(\left(\frac{T'}{T}+2\alpha T\right)x+\bar{\bar{T}}\right)y,
\end{equation*}
where
\begin{equation}\label{eq:25}
  \xi=x+\frac{1}{2}\left(\frac{T'}{T}+\alpha T\right)y^2 -
    \bar{T}y.
\end{equation}
Substituting to Equation~\eqref{eq:4}, one obtains
\begin{equation*}
  \boxed{\left(\left(\frac{T'}{T}+2\alpha T\right)\xi +\bar{T}^2 +
      \bar{\bar{T}}\right)\Phi_{\xi\xi}-\Phi_{\xi t} - \alpha T\Phi_\xi
    +\bar{T}' +2\alpha T\bar{T}=0.}
\end{equation*}
Of course, the equation can be solved explicitly, but the final result is too
cumbersome: it is easily shown that
\begin{equation*}
  \Phi_\xi=\left(\bar{Z}\left(\xi e^{\int a\,dt}-\int be^{\int a\,dt}\,dt\right)
    +\int ce^{\alpha\int 
      T\,dt}\,dt\right)e^{-\alpha\int T\,dt},
\end{equation*}
where~$\bar{Z}$ is an arbitrary function in one variable and
\begin{equation*}
  a=\frac{T'}{T}+2\alpha T,\quad b=\bar{T}^2+\bar{\bar{T}},\quad
  c=\bar{T}'+2\alpha T\bar{T}.
\end{equation*}
Thus,
\begin{equation*}
  \Phi=Z\left(\xi e^{\int a\,dt}-\int be^{\int a\,dt}\,dt\right)e^{-2\alpha\int T\,dt}
    +\left(\int ce^{\alpha\int 
      T\,dt}\,dt\right)\xi e^{-\alpha\int T\,dt} + \phi(t),
\end{equation*}
and the corresponding family of solutions is\par\medskip
\fbox{%
\parbox{5.5cm}{%
\begin{multline*}
  u=Z\left(\xi e^{\int a\,dt}-\int be^{\int a\,dt}\,dt\right)e^{-2\alpha\int T\,dt}
    +\left(\int ce^{\alpha\int 
      T\,dt}\,dt\right)\xi e^{-\alpha\int T\,dt} + \phi(t)\\
  -\left(\frac{1}{6}\frac{T''}{T} + \alpha T' +
    \frac{2}{3}\alpha^2T^2\right)y^3
  + \frac{1}{2}\left(\bar{T}'+2\alpha
    T\bar{T}\right)y^2
  -  \left(\left(\frac{T'}{T}+2\alpha T\right)x+\bar{\bar{T}}\right)y
\end{multline*}}%
}\medskip\par\noindent
with~$\xi$ given by~\eqref{eq:25}.

\paragraph{\underline{Subcase~$01$}}
\label{sec:subcase-01}
Since~$\beta\neq 0$, we can set~$\beta=-1$ and the defining equations become
\begin{equation}\label{eq:24}
  \frac{dx}{2 x + (T_5'-\alpha) y + T_4}=\frac{dy}{y+T_5}
  =\frac{dt}{0}= \frac{du}{3u+(T_5'-2\alpha)x+T_4'y+\frac{1}{2}T_5''y^2- T_6}.
\end{equation}
Let us introduce the notation~$T=T_5$, $\bar{T}=T_4-T_5(T_5'-\alpha)$. Then
the integrals of \eqref{eq:24} are
\begin{multline*}
  t=\const,\qquad \frac{x+\frac{1}{2}\bar{T}}{(y+T)^2} + \frac{T'-\alpha}{y+T}
  = \const\\
  \frac{u+(x+\frac{1}{2}\bar{T})(T'-2\alpha)+\frac{1}{3}\bar{\bar{T}}}{(y+T)^3}
  + \frac{(T'-\alpha)^2+\frac{1}{2}\bar{T}'}{(y+T)^2}
  +\frac{1}{2}\frac{T''}{y+T}=\const,  
\end{multline*}
where
\begin{equation*}
  \bar{\bar{T}}=-\frac{1}{2}(T'-2\alpha)\bar{T} -
  \big(\bar{T}'+T'(T'-\alpha)+TT''\big)T + \frac{1}{2}T^2T''-T_6.
\end{equation*}
Consequently, the general solution is
\begin{equation*}
  \Psi\left(\frac{u+(x+\frac{1}{2}\bar{T})(T'-2\alpha)
      +\frac{1}{3}\bar{\bar{T}}}{(y+T)^3} +
    \frac{(T'-\alpha)^2+\frac{1}{2}\bar{T}'}{(y+T)^2}
    +\frac{1}{2}\frac{T''}{y+T},\
    \frac{x+\frac{1}{2}\bar{T}}{(y+T)^2} + \frac{T'-\alpha}{y+T},\ t\right)=0,
\end{equation*}
or
\begin{equation*}
  u=(y+T)^3\Phi(\xi,t) - \frac{1}{2}T''(y+T)^2 - \left((T'-\alpha)^2 +
    \frac{1}{2}\bar{T}'\right)(y+T) -
  (T'-2\alpha)\left(x+\frac{1}{2}\bar{T}\right) - \frac{1}{3}\bar{\bar{T}},
\end{equation*}
where
\begin{equation*}
  \xi=\frac{x+\frac{1}{2}\bar{T}}{(y+T)^2} + \frac{T'-\alpha}{y+T}.
\end{equation*}
Substituting to Equation~\eqref{eq:4}, one obtains
\begin{equation*}
  \boxed{(4\xi^2-3\Phi)\Phi_{\xi\xi}-\Phi_{\xi t} -6\xi\Phi_\xi + \Phi_\xi^2 +
    6\Phi=0.}
\end{equation*}

\subsubsection{Case~$1$}
\label{sec:case-1}
The defining equation is now
\begin{multline*}
  \frac{dx}{(T_3'-2\beta) x + (T_5'-\alpha) y +\frac{1}{2}T_3''y^2 + T_4}
  =\frac{dy}{(T_3'-\beta) y+T_5} =\frac{dt}{T_3}\\= \frac{du}{(T_3' - 3\beta)u
    + (T_5'-2\alpha) x + T_4'y+ T_3''xy + \frac{1}{2}T_5''y^2 +
    \frac{1}{6}T_3'''y^3 - T_6}
\end{multline*}
and since~$T_3\neq 0$ we may set~$T'=1/T_3$. The integrals are
\begin{align*}
  &T'ye^{\beta T}-\bar{T}=\const,\\
  &T'xe^{2\beta T}+\frac{1}{2}T_3'\xi^2 - \left(\left(\frac{\bar{T}}{T'}\right)' -
      \beta \bar{T}'-\alpha k(\beta)\right)\xi-\bar{\bar{T}}=\const,\\
  &T'ue^{3\beta T}-T_{00}-T_{10}\xi -T_{01}\eta - T_{20}\xi^2 - T_{11}\xi\eta
  - T_{30}\xi^3=\const,
\end{align*}
where
\begin{align*}
  \xi&=T'ye^{\beta T}-\bar{T},\\
  \eta&=T'xe^{2\beta T}+\frac{1}{2}T_3'\xi^2 -
  \left(\left(\frac{\bar{T}}{T'}\right)' -
    \beta \bar{T}'-\alpha k(\beta)\right)\xi-\bar{\bar{T}},\\
  \bar{T}&=\int T_5(T')^2e^{\beta T}\,dt,\\
  \bar{\bar{T}}&=\int\left(T_4(T')^2e^{2\beta T} + (T_5'-\alpha)\bar{T}T'
    e^{\beta T} + \frac{1}{2}T_3''(\bar{T})^2\right)\,dt,\\
  T_{00}&=\int\left(T_3''\bar{\bar{T}}\bar{T} + \frac{T_3'''\bar{T}^3}{6T'} +
    \left(T_4'\bar{T}T' +(T_5'-2\alpha)T' +
      \frac{1}{2T_5''(\bar{T})^2}\right)e^{\beta T}
    - T_6(T')^2e^{3\beta T}\right)\,dt,\\
  T_{10}&=\int\left(\frac{T_3'''(\bar{T})^2}{2T'} + T_3''\left( \bar{\bar{T}}
      +\bar{T}\left( \left(\frac{\bar{T}}{T'}\right)'-\alpha k(\beta) -
        \beta\bar{T} \right) \right)\right.\\ &\left.+\left(T_5''\bar{T} +
      (T_5'-2\alpha)\left( \left(\frac{\bar{T}}{T'}\right)'-\alpha k(\beta) -
        \beta\bar{T} \right)\right)e^{\beta T} + T_4'T'e^{2\beta T}
  \right)\,dt,\\
  T_{01}&=\int\left(T_3'''\bar{T}+(T_5'-2\alpha)T'e^{\beta T}\right)\,dt,\\
  T_{20}&=\int\left(T_3''\left(\frac{\bar{T}}{2T'} +
      \left(\frac{\bar{T}}{T'}\right)'-\alpha k(\beta) -
      \beta\bar{T} -\frac{1}{2}T_3'\bar{T}\right) + \frac{1}{2}\left(T_5'' -
    (T_5'-2\alpha)T'\right)e^{\beta T}\right)\,dt,\\ 
  T_{11}&=\int T_3''\,dt=T_3',\\
  T_{30}&=\int\left(\frac{T_3'''}{6T'}-\frac{1}{2}T_3''T_3'\right)\,dt =
  \frac{1}{6}T_3''T_3 -\frac{1}{3}(T_3')^2,
  \intertext{and} k(\beta)&=\int T'e^{\beta
    T}\,dt=
  \begin{cases}
    \dfrac{e^{\beta T}}{\beta},&\beta\neq 0,\\
    T,&\beta=0.
  \end{cases}
\end{align*}
Thus, the general solution is
\begin{equation}
  \label{eq:26}
  u=\left(\frac{\Phi(\xi,\eta)+T_{00}+T_{10}\xi +T_{01}\eta + T_{20}\xi^2 +
      T_{11}\xi\eta 
      + T_{30}\xi^3}{T'}\right)e^{3\beta T}.
\end{equation}
Substituting~\eqref{eq:26} to Equation~\eqref{eq:4}, one obtains
\begin{equation*}
  \Phi_{\xi\xi}=\left(\beta\xi-\Phi_\eta\right)\Phi_{\xi\eta} +
  \left(2\beta\eta+\alpha\kappa+\Phi_\xi\right)\Phi_{\eta\eta} -
  \beta\Phi_\kappa -2\alpha\kappa, 
\end{equation*}
where
\begin{equation*}
  \kappa=e^{\beta T}-\beta k(\beta),
\end{equation*}
i.e.,
\begin{equation*}
  \kappa=
  \begin{cases}
    0,&\beta\neq 0,\\
    \xi,&\beta=0.
  \end{cases}
\end{equation*}
Thus, the reductions are
\begin{equation*}
  \boxed{\Phi_{\xi\xi}=(\beta\xi-\Phi_\eta)\Phi_{\xi\eta} +
    (2\beta\eta+\Phi_\xi)\Phi_{\eta\eta} -\beta\Phi_\eta=0}
\end{equation*}
for~$\beta\neq 0$ and
\begin{equation}\label{eq:27}
  \boxed{\Phi_{\xi\xi}=(\alpha\xi+\Phi_\xi)\Phi_{\eta\eta} -
    \Phi_\eta\Phi_{\xi\eta}-2\alpha} 
\end{equation}
for~$\beta=0$. Note that in the case~$\alpha\neq 0$ Equation~\eqref{eq:27}
transforms to the Gibbons-Tsarev equation (see~\cite{GT})
\begin{equation*}
  \boxed{\Phi_{\xi\xi}=\Phi_\xi\Phi_{\eta\eta}-\Phi_\eta\Phi_{\xi\eta}-\alpha}
\end{equation*}
by~$\Phi\mapsto\Phi-\alpha\xi^2/2$.

\section{Summary of results}
\label{sec:summary-results}
Below, a concise exposition of the obtained results is given\footnote{We use
  the notation~$\infty^{k\cdot \tau}$ to indicate the infinite-dimensional
  component corresponding to~$k$ arbitrary functions in~$\tau$} in
Table~\ref{tab:summary} on~p.~\pageref{tab:summary}.
\begin{table}[htb]
  \centering
  \begin{tabular}{l|l|l|l}\hline
    Eqn&$\dim(\sym\mathcal{E})$&Reductions&Comments\\\hline
    \eqref{eq:1}&$1+\infty^{2\cdot x}+\infty^{2\cdot
      z}$&$X\Phi_{xz}-X'\Phi_z=0$&Solves explicitly\\
    &&$2\Phi=\Phi\Phi_{xz}-\Phi_x\Phi_z$&Transforms\\ &&& to the Liouville
    equation\\
    &&$\Phi_{\xi\xi}=X'\Phi_\xi-X\Phi_{x\xi}$&Solves explicitly\\
    &&$\Phi_{\xi\xi}=\Phi_x\Phi_\xi-\Phi\Phi_{x\xi}$&\\
    &&$(1+Z\Phi_z)\Phi_{\xi\xi}=Z\Phi_\xi\Phi_{\xi z}+Z'\Phi_\xi^2$&Solves
    explicitly\\ 
    &&$(1+\xi\Phi_z)\Phi_{\xi\xi}-\xi\Phi_\xi\Phi_{\xi z}+\Phi_\xi\Phi_z=0$&\\
    &&$\Phi_{\xi\xi}=\Phi_\xi\Phi_{\eta\eta}-\Phi_\eta\Phi_{\xi\eta}$&Linearizes
    by the\\
    &&&Legendre transformation\\
    &&$\Phi_\eta\Phi_{\xi\eta}-\Phi_\xi\Phi_{\eta\eta}=e^\eta\Phi_{\xi\xi}$&\\\hline
    \eqref{eq:2}&$1+\infty^{1\cdot y}+\infty^{3\cdot
      t}$&$\Phi_{yt}=T\Phi_y$&Solves explicitly\\
    &&$\Phi_{yt}=2\Phi\Phi_y$&Reduces\\
    &&&to the Riccati equation\\
    &&$(\alpha\xi+\Phi_\xi)\Phi_{\xi
      y}-\Phi_y(\Phi_{\xi\xi}+2\alpha)=0$&Solves explicitly for~$\alpha=0$\\
    &&$T\Phi_{xx}=T'$&Solves explicitly\\
    &&$\left(\frac{T'}{T}\xi+\bar{T}\right)\Phi_{\xi\xi} + \Phi_{\xi
      t}=0$&Solves explicitly\\
    &&$\Phi_{\xi t}=4\Phi\Phi_\xi-\xi\Phi_\xi^2+2\xi\Phi\Phi_{\xi\xi}$&\\
    &&$\Phi_{\eta\eta}+(\alpha\xi+\Phi_\eta)\Phi_{\xi\eta} = \Phi_\eta(2\alpha
    +\Phi_{\xi\xi})$&Linearizes
    by the \\
    &&&for~$\alpha=0$\\
    \hline
    \eqref{eq:3}&$\infty^{2\cdot x}+\infty^{1\cdot y}+\infty^{1\cdot
      t}$&$\Phi_{yt}=0 $&Solves explicitly\\
    &&$\bar{X}\Phi_{xt}-\bar{X}'\Phi_t=0$&Solves explicitly\\
    &&$\bar{X}\Phi_{xy}-\bar{X}'\Phi_y=0$&Solves explicitly\\
    &&$\Phi_{\xi\xi}=\bar{X}\Phi_{x\xi}-\bar{X}'\Phi_\xi$&Solves explicitly\\
    &&$(1+\Phi_\xi)\Phi_{y\xi}=\Phi_y\Phi_{\xi\xi}$&Solves explicitly\\
    &&$(1+\Phi_\xi)\Phi_{t\xi}=\Phi_t\Phi_{\xi\xi}$&Solves explicitly\\
    &&$\Phi_\eta\Phi_{\xi\xi}+
    (\Phi_\eta-\Phi_\xi-1)\Phi_{\xi\eta}-\Phi_\xi\Phi_{\eta\eta}=0$&Linearizes
    by the \\ 
    &&&Legendre transformation\\   \hline
    \eqref{eq:4}&$2+\infty^{4\cdot t}$&$\Phi_{yy}=
    (T'-T^2)y+T\bar{T}-\bar{T}'$&Solves explicitly\\
    &&$\Phi_{yy}=\frac{2\Phi_y-T'}{y+T}+T'' +\frac{\bar{T}^2}{(y+T)^3}$&Solves
    explicitly\\
    &&$\left(\left(\frac{T'}{T}+2\alpha T\right)\xi +\bar{T}^2+
      \bar{\bar{T}}\right)\Phi_{\xi\xi}$&Solves explicitly\\
    &&$ -\Phi_{\xi t} - \alpha T\Phi_\xi +
    \bar{T}' +2\alpha T\bar{T}=0$&\\
    &&$(4\xi^2-3\Phi)\Phi_{\xi\xi}-\Phi_{\xi t}- 6\xi\Phi_\xi+\Phi_\xi^2+6\Phi=0$&\\
    &&$\Phi_{\xi\xi}=(\beta\xi-\Phi_\eta)\Phi_{\xi\eta} +
    (2\beta\eta+\Phi_\xi)\Phi_{\eta\eta} -\beta\Phi_\eta=0$&Linearizes by
    the\\
    &&&Legendre transformation\\
    &&&for~$\beta=0$\\
    &&$\Phi_{\xi\xi}=(\alpha\xi+\Phi_\xi)\Phi_{\eta\eta}-\Phi_\eta\Phi_{\xi\eta}
    - 2\alpha$&Reduces to the\\
    &&&Gibbons-Tsarev equation,\\
    &&&for~$\alpha\neq 0$\\
    &&&Linearizes by the \\
    &&&Legendre transformation\\
    &&&for~$\alpha=0$\\\hline  
  \end{tabular}\smallskip
  \caption{Summary of reductions}
  \label{tab:summary}
\end{table}

\section*{Acknowledgements}
\label{sec:acknowledgements}

The authors are grateful to E.~Ferapontov, M.~Marvan, and A.~Sergyeyev for
discussions. Computations of symmetry algebras were fulfiled using the
\textsc{Jets} sofware,~\cite{Jets}.

\end{document}